\documentclass[reprint,
nofootinbib, 
amsmath,
amssymb,
aps,
prd,
superscriptaddress]{revtex4-2}

\usepackage{graphicx}
\usepackage{dcolumn}
\usepackage{bm}
\usepackage{siunitx}

\usepackage{makecell}
\usepackage{multirow}
\usepackage{tabularx}
\usepackage{hyperref}
\usepackage{aas_macros}
\usepackage{xcolor}

\newcommand{\ud}{\mathrm{d}}
\newcommand{\msun}{\,\mathrm{M}_\odot}
\newcommand{\REV}[1]{{\color{black}#1}}

\begin{document}

\preprint{}

\title{The deci-Hz gravitational wave signal from the collapse of rotating very massive stars}

\author{Bailey Sykes}
\email{bailey.sykes@monash.edu}
\affiliation{School of Physics and Astronomy, Monash University, Clayton, VIC, 3800, Australia}

\author{Jade Powell}
\email{dr.jade.powell@gmail.com}
\affiliation{Centre for Astrophysics and Supercomputing, Swinburne University of Technology, Hawthorn, VIC 3122, Australia}

\author{Bernhard M\"{u}ller}
\email{bernhard.mueller@monash.edu}
\affiliation{School of Physics and Astronomy, Monash University, Clayton, VIC, 3800, Australia}

\author{Alexander Heger}
\email{alexander.heger@monash.edu}
\affiliation{School of Physics and Astronomy, Monash University, Clayton, VIC, 3800, Australia}


\begin{abstract}

We calculate the gravitational wave signal from the collapse of a rotating $300\,\msun$ star just above the upper end of the pair-instability supernova regime. 
The large-scale asymmetries that develop during the collapse produce a strong signal in the deci-Hz range that has a characteristic shape that is likely amenable to a template-based search. Although tempered by the low incidence of nearby astrophysical candidates, the most ambitious designs for deci-Hz detectors could detect such signals out to distances of $\sim 250\,\mathrm{Mpc}$, possibly at a rate of $0.3\,\mathrm{yr}^{-1}$. Additional numerical modeling is required to refine these rate estimates, however. We also demonstrate that proposed deci-Hz GW observation missions, even without confident single-source detections, could be used to provide information about the population of astrophysical candidates and, in particular, constrain the initial mass function of very massive stars.

\end{abstract}

\maketitle

\section{Introduction}
Gravitational wave (GW) observations of merging black holes and neutron stars have furnished revolutionary insights into the mass distribution of these compact objects, constraining their formation by stellar collapse \citep{abbott_23}. Further challenges and opportunities for GW astronomy include the detection of
other sources such as core-collapse supernovae (CCSNe) \citep{kalogera_21} and the extension of observations 
beyond the current frequency range of ${\sim} 10\texttt{-}10^{3}\,\mathrm{Hz}$.

Efforts to detect GWs from stellar collapse \cite{Abbott_et_al:2020, Abac_et_al:2025} have historically focused on the frequency range covered by
LIGO \citep{LIGO:2015}, Virgo \citep{VIRGO:2015} and KAGRA \citep{KAGRA:2021} and their prospective third-generation
successors \cite{punturo_10,lisa_17,evans_21}. However, the deci-Hz range of $\mathord{\sim}10^{-2}\texttt{-}10^{1}\,\mathrm{Hz}$, targeted by the proposed
space-based detectors DECIGO \citep{kawamura_11}, BBO \citep{harry_06} and IndIGO-D \citep{Sharma_et_al:2026}, is emerging as another interesting window for stellar collapse. CCSNe may radiate in this band due to anisotropic neutrino emission and asymmetric shock expansion \citep{epstein_78, murphy_09,marek_09,Vartanyan_Burrows:2020,Richardson_et_al:2022,choi_24,powell_24}. Anisotropic neutrino emission is also relevant for supermassive stars collapsing via the general relativistic instability \citep{Li_Fuller_Kishimoto:2018}.
In addition, the collapse of rotating supermassive stars could provide strong signals extending into the deci-Hz band \citep{Saijo_et_al:2002,montero_12,shibata_16,uchida_17}, especially if it entails the formation of disks subject to non-axisymmetric instabilities \citep{Saijo_Hawke:2009, uchida_17}.

We here investigate the black-hole forming collapse of rotating \emph{very massive stars} of initially several $100\,\mathrm{M}_\odot$ as a source of GWs. Stars with initial masses of $\gtrsim 140\,\mathrm{M}_\odot$ undergo collapse driven by pair instability \citep{Heger_et_al:2003} after carbon burning. Below  ${\sim} 260\,\mathrm{M}_\odot$, this is expected to lead to the complete disruption of the star by violent oxygen and silicon burning. At higher masses, black hole formation ensues and may be accompanied by the formation of an accretion disk \citep{Fryer2001,uchida_19,shibata_26}. 

The prevalence and fate of these very massive stars constitutes a major unresolved problem in astrophysics. GW observations have now revealed  stellar mass black holes of up to
${\sim} 140\msun$ in the merger event GW231123 \citep{abac_25}, and it is critical to ascertain whether such objects are formed directly by stellar collapse or by hierarchical mergers. The formation of such massive black holes by stellar collapse of Population~III stars at high redshift is one of the scenarios for seeding intermediate-mass black holes \citep{Madau_Rees:2001} and jump-starting a process of accretion and merging that eventually produces supermassive black holes. Finally, the rate and mass range for
pair instability supernovae (PISNe) and their demarcation from the regime of complete collapse at higher masses
are major challenges for stellar astrophysics and transient astronomy, making this mass range a key target for current and upcoming transient surveys \citep{Moriya_et_al:2022}.

Although evidence for the explosion or black-hole formation of very massive stars is still ambiguous, observations and theory support the plausibility of this scenario.
Stellar birth masses well above $100\msun$ are well attested even in the Milky Way and its satellites \citep{figer_02,schneider_18},
and mass loss may be sufficiently attenuated at low metallicity for these to evolve towards pair instability. Conditions for very massive star formation may be more favorable for Population III stars if the early Universe initial mass function (IMF) was top-heavy \citep{Stacy_Bromm_Lee:2016}.
Possible candidates for PISNe have been proposed among
observed transients \citep{Schulze:2024}, and abundances in the very-metal poor star J1010+2358
have been interpreted as indicative of pollution by PISN ejecta \citep{xing_23}.

Previous studies of GWs from the collapse of very massive stars have primarily considered GW emission by triaxial instabilities in accretion disks \citep{uchida_19,shibata_26}, with early estimates of the strain amplitude of $h\sim 10^{-21}$ for events at distance of order $1\,\mathrm{Gpc}$. More recent studies based on general relativistic simulations also predict
a strong signal with characteristic frequencies of tens of Hz and a
detection horizon of up to redshift $z\sim 1$ with third-generation ground-based detectors \citep{shibata_26}. GWs from asymmetric neutrino emission has, to our knowledge, not been studied for the population of very massive stars. It may produce a GW signal in the deci-Hz range, following the behavior of both massive \citep{Vartanyan_Burrows:2020} and supermassive stars \citep{Li_Fuller_Kishimoto:2018}.

We here demonstrate that axisymmetric collapse of matter alone already produces a strong GW signal in the deci-Hz range. Our study extends previous work that was either conducted in the Newtonian approximation \citep{Fryer2001} or in general relativity, but with a somewhat idealized progenitor structure and without nuclear burning \citep{shibata_26}, to predict GW amplitudes well beyond black-hole formation. Our results highlight the added potential of the deci-Hz band as a probe of the evolution and collapse of very massive stars, and the origin of the most massive stellar black holes.

This paper is organized as follows. In Section~\ref{sec:methods} we describe the very massive stellar progenitor we use, and the configuration of our hydrodynamics code. This is followed in Section~\ref{sec:results} by an overview of the GW signal. We address the detectability of this signal in Section~\ref{sec:detect} before offering concluding remarks in Section~\ref{sec:conclusions}.

\section{Methods}
\label{sec:methods}
\subsection{Initial model}

We simulate the collapse of a rotating, zero-metallicity star of $300 \msun$ (with a $180 \msun$ helium core) that was previously studied by \citet{Fryer2001}. Starting from rigid rotation
on the zero-age main sequence with 20\% of the Keplerian velocity at the surface, the rotational evolution is treated according to \citet{Heger_Langer_Woosley:2000}, accounting for various hydrodynamic instabilities, but neglecting magnetic fields. The initial rotation is comparable to typical observed O stars. This it not an extreme choice as there are plausible scenarios for obtaining and maintaining rapid rotation in very massive stars, e.g., by chemically homogeneous evolution in binary systems \citep{Popa_deMink:2025}.
The model is already collapsing due to pair instability and has photo-disintegrated to a mix of alpha particles, protons and neutrons in the innermost ${\sim} 30 \, \msun$ when it is mapped from \textsc{Kepler} \citep{Weaver_Zimmerman_Woosley:1978} to the general relativistic supernova simulation code \textsc{CoCoNuT-FMT}.

\subsection{Simulation setup}

The subsequent collapse of the star is then simulated in axisymmetry (2D) with \textsc{CoCoNuT-FMT}. The code solves the equations of relativistic hydrodynamics and the metric equations in the conformally flat approximation
\citep{Isenberg:2008} combined with an excision scheme for black-hole space-times \citep{CorderoCarrion_et_al:2014,Sykes_Mueller:2025}. The metric is treated as spherically symmetric in this study.
We also include neutrino transport with the fast multi-group transport (FMT) scheme of \citet{Mueller_Janka:2015} and a 19-species nuclear network \citep{Weaver_Zimmerman_Woosley:1978, Timmes:1999}. The SFHo equation of state \citep{Steiner_Hempel_Fischer:2013} is used in the high-density regime. The grid resolution is $N_\mathrm{r} \times N_\theta = 600 \times 128$, for radial and polar directions respectively. The outer boundary is placed at $10^{11} \, \mathrm{cm}$.

GW amplitudes $A^{\mathrm{E2}}_{20}$
\citep{Thorne:1980} are extracted using a version of the time-integrated quadrupole formula with non-linear corrections for strong-field gravity \citep{finn_90,blanchet_90,Mueller_Janka_Marek:2013},
\begin{align}
    \label{eqn:aetwo}
    A^{\mathrm{E2}}_{20} = & \frac{16 \pi^{3/2} G}{\sqrt{15} c^{4}} \frac{\partial}{\partial t} \iint \phi^{3} r^{3} \sin \theta  \sqrt{\alpha(\alpha / \phi^{2} - \beta^{r} )}
    \\ \nonumber
        & \bigg( \hat{S}_{r} (3 \cos^{2} \theta - 1) - 3 r^{-1} \hat{S}_{\theta} \sin \theta \cos \theta \bigg)
        \ud \theta \ud r,
\end{align}
where $\hat{S}_{i} = \sqrt{\gamma} S_{i}$ are components of the densitized covariant three-momentum of the fluid \citep{Mueller_Janka_Dimmelmeier:2010}. Compared to \citep{Mueller_Janka_Marek:2013}, we introduce
a redshift scaling term, $\phi^{3} \sqrt{\alpha(\alpha / \phi^{2} - \beta^{r} )}$
in Equation~(\ref{eqn:aetwo}), which approximates the redshifting of metric perturbations sourced close to the apparent horizon by appealing to a relativistic conservation law of the GW amplitude \citep{Laguna_et_al:2010}, analogous to that for electromagnetic waves \citep{gravitation_mtw}. The formula reduces to the form of \citet{Mueller_Janka_Marek:2013} under their original assumptions, i.e., for vanishing shift and $\alpha\phi^2=1$.
To ascertain that GW extraction with a modified quadrupole is appropriate, we verified that
the region close to the horizon does not contribute appreciably to the low-frequency signal, which arises primarily from motions outside a radius of $10^{3}\,\mathrm{km}$ (see Figure~\ref{fig:amplitude}).

\REV{We note that Equation \eqref{eqn:aetwo} corrects two typographic errors in \citet{Mueller_Janka_Marek:2013} and prior works. Namely, the pre-factor of $16$ was incorrectly written as $32$, and the $\theta$ momentum term included a sign error. Previous works are largely unaffected by this error, however, as the formula is implemented correctly in the code.}

\section{Results}
\label{sec:results}
The collapse of the core is initially driven by photo-disintgeration and later accelerated by deleptonisation; consequently, prompt black-hole formation occurs after $0.9\,\mathrm{s}$. Different from \citet{Fryer2001}, there is no transient, hypermassive proto-neutron star. Within a second, the black hole grows above $100\msun$ and eventually swallows most of the star. Due to the interaction of centrifugal forces and nuclear burning, an accretion disk with a complex and evolving morphology is formed after several seconds, and some material is eventually ejected. A detailed discussion of the collapse dynamics of this model, including its other multi-messenger signals, is the subject of a future work.

\begin{figure}
    \centering
    \includegraphics[width=\linewidth]{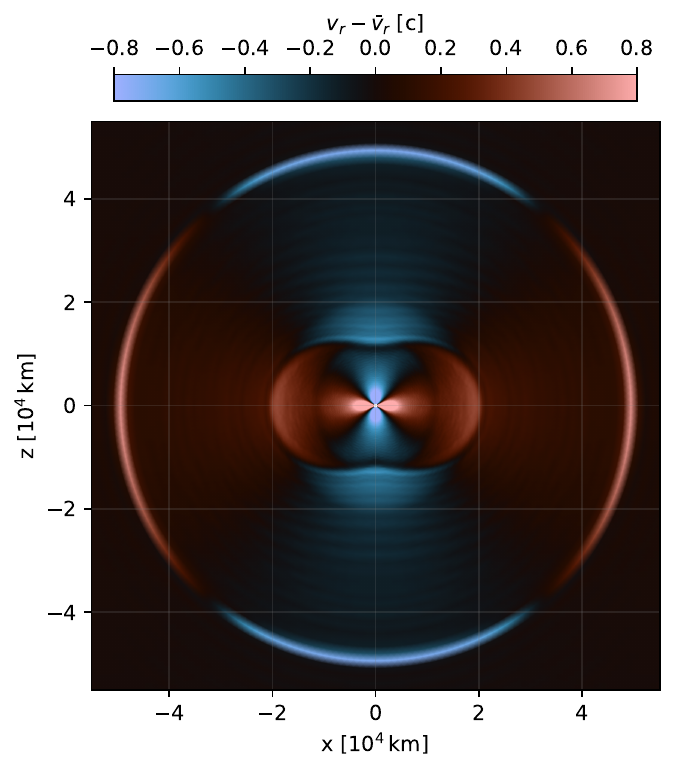}
    \caption{Slice of the local deviation of radial velocity from the spherical average, mirrored across the rotation axis (vertical/z axis) at a simulation time of $2.2 \, \mathrm{s}$ (${\sim}1.3 \, \mathrm{s}$ after BH formation). Blue tones indicate material falling in faster than the average, and red tones indicate slower infall.}
    \label{fig:slice}
\end{figure}

\begin{figure*}
    \centering
    \includegraphics[width=\linewidth]{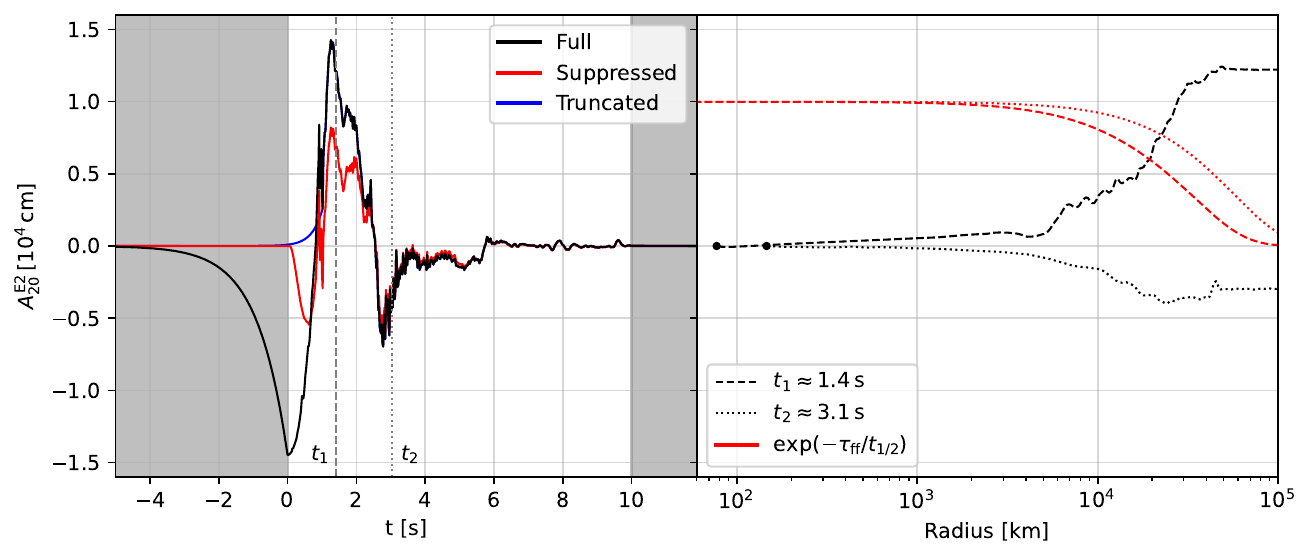}
    \caption{\textbf{Left}: GW amplitude, $A^{\mathrm{E2}}_{20}$, over the entire simulation duration ($0-10 \, \mathrm{s}$). Small sections of the padded signal (shaded regions) are also shown, in which the tapering of the amplitude is evident. The black line shows the full signal produced by directly applying Equation~\eqref{eqn:aetwo} to the simulation data. The red line shows the amplitude if initial low-frequency contributions are suppressed by a time-dependent exponential factor (see main text for details). The blue line,  which is identical to the black after $1 \, \mathrm{s}$, shows the tapering of the signal if the early-time negative amplitude part is discarded.
    The amplitudes are downsampled to $50 \, \mathrm{Hz}$ to omit high-frequency noise; this also avoids sampling a (post-processing) glitch from the time derivative in Equation \eqref{eqn:aetwo} due to BH excision switching on in the code. 
    \textbf{Right}: Cumulative radial integrations of $A^{\mathrm{E2}}_{20}$ (the full signal, not the suppressed or truncated versions) at the two times indicated by vertical lines on the LHS. The radius of the BH's apparent horizon is denoted by a black dot. For reference, we also show the radial dependence of the suppression factor at $t_1$ and $t_2$ used in calculating the suppressed amplitude signal in the LHS panel.}
    \label{fig:amplitude}
\end{figure*}

The time-dependent global asymmetries produced from the rotational collapse are shown in Figure~\ref{fig:slice}. These asymmetries, on scales ranging from ${\sim} 10^{3} \, \mathrm{km}$ up to several $10^{4} \, \mathrm{km}$, produce a characteristic low-frequency GW signal, shown in Figure~\ref{fig:amplitude} as a black line, with one prominent crest reaching \REV{${\sim} 13\mathord,000\,\mathrm{cm}$} in distance-normalized amplitude, followed by a weaker trough slightly exceeding \REV{$6\mathord,000 \, \mathrm{cm}$}. \REV{The large negative amplitude near $t=0 \, \mathrm{s}$ in the full signal, however, is not a reliable component of the signal, instead stemming from non-stationarity of the initial model when rotation is added. While we delay a more detailed discussion of this feature, we note here that Figure~\ref{fig:amplitude} introduces two modified waveforms, labeled ``suppressed'' and ``truncated'', which attempt to lessen the non-physical component of the waveform (see Section~\ref{sec:truncation} for details). The blue ``truncated'' waveform is likely the more realistic scenario as it best preserves the prominent crest and subsequent trough, which are the robust features of the waveform.}

\REV{The signal morphology is similar to the first crest and trough of the bounce signal during rotational iron core collapse \citep{Dimmelmeier_Ott_Marek_Janka:2008, Fuller_et_al:2015}, and indeed, the rise phase of the crest is produced by the collapse of the rotationally deformed core. A key difference, however, is the absence of a core bounce reversing the infall of material. Instead, the trough and the return of the amplitude to zero are best explained by a redistribution of momentum into and out of the quadrupole in the polar and equatorial regions. Whereas the radial momenta dominate Equation~(\ref{eqn:aetwo}), this redistribution of momentum is mediated by meridional flows, which drain momentum from the polar region to the equator. Accelerated accretion onto the nascent BH along the poles also directly drains a component of the quadrupole.
Similar to the GWs during the bounce phase of iron core collapse, the characteristic zig-zag waveform -- potentially with a bias towards the positive amplitude crest -- is likely robust to moderate variations in progenitor structure and rotation \citep{Richers_et_al:2017}. Additional simulations will be required to fully confirm this however.}

In addition to the low-frequency signal, the waveform in Figure~\ref{fig:amplitude} also has a high-frequency component which is strongest immediately after a stable, rotationally-supported disk forms at around $t=3\, \mathrm{s}$. There is also a brief high-frequency burst accompanying black-hole formation. These features may be less robust than the prominent crest and trough, but are of secondary relevance for deci-Hz observations. \REV{As the disk does not form until about $3 \, \mathrm{s}$ after the start of our simulation, well after the beginning of the low-frequency waveform, the disk dynamics and the disk instabilities cannot contribute to the deci-Hz GW signal. Likewise, nuclear burning does not notably impact the dynamics generating the low-frequency signal. This is evident from the black lines on the RHS of Figure~\ref{fig:amplitude}, which show the cumulative integral (from the interior outwards) of Equation~\eqref{eqn:aetwo}. The vast majority of the total amplitude is sourced from $r\gtrsim10^{4} \, \mathrm{km}$ -- well above the disk-forming region, or any regions with active nuclear burning. Additionally, that the signal is sourced from such a large radius alleviates concerns that the approximate redshift correction in Equation~\eqref{eqn:aetwo} may impact the waveform negatively; so far from the black-hole, the space-time is relatively weakly curved and the correction factor is near to one. This conveniently also sidesteps the issue of the accuracy of Equation~\eqref{eqn:aetwo} in the vicinity of a rotating black-hole using a CFC metric. Although the CFC approximation has been shown to be reasonably accurate even for rotating space-times where it is formally not suitable \citep{Cerda-Duran_et_al:2005}, the low-frequency signal originates from a region where general relativistic effects can largely be ignored; thus, the quality of the approximation is very high.}

Unlike high-frequency GW signals from CCSNe and disk instabilities, a deci-Hz signal from large-scale rotational asymmetries is not impacted by the differences in turbulence between 2D and 3D simulations \citep{Couch_OConnor:2014}. Overall, our assumption of axisymmetry likely has minimal effect on the low-frequency component of the signal, however 3D simulations will be needed to rigorously confirm this because, e.g., \REV{post-processed instability criteria can provide an incomplete understanding of the growth of instabilities \citep{Shibata_Sekiguchi:2005, Dimmelmeier_Ott_Marek_Janka:2008}. }Non-axisymmetric effects, and, in particular, additional GW signatures due to triaxial instabilities, will be investigated in a future work.

\subsection{Correction of non-hydrostatic initial conditions}
\label{sec:truncation}
Returning now to the large negative amplitudes of the full signal in Figure~\ref{fig:amplitude}, we note that such a large amplitude emerging immediately at the start of the simulation is not physical. At such an early phase of the collapse, the star should not be strongly deformed compared to its late-phase evolution. Instead, it is the relaxation of the progenitor -- which is modeled in spherical symmetry -- after imposing a rotation profile and mapping to 2D which causes the initial strong component of the GW signal. This process breaks the near-hydrostaticity of the star by introducing new centripetal forces which immediately act on the momentum terms in Equation~\eqref{eqn:aetwo}, primarily at large ($r\gtrsim 3 \times 10^{4} \, \mathrm{km}$) radii. On the other hand, the later phase of the signal is clearly physical; the changing geometry of the infalling material must produce a GW signal of the shape described in the previous section.

The separation of this artificial relaxation signal and the real GW waveform is not straightforward since they come from nearby regions of the star. 
While the optimal solution is to deform the progenitor into a hydrostatic rotational equilibrium before running the simulation, such a task is beyond the scope of this work. Instead we present some approximate methods to recover the real signal with the red and blue lines in Figure~\ref{fig:amplitude}. 

For the red line, we suppress the integrand in Equation~\eqref{eqn:aetwo} by a radius and time-dependent exponential, i.e.,

\begin{align}
    \label{eqn:aetwo_suppressed}
    A^{\mathrm{E2}}_{20} = & \frac{16 \pi^{3/2} G}{\sqrt{15} c^{4}} \frac{\partial}{\partial t} \iint \mathcal{I}(r,\theta,t) \exp\left( \frac{-\tau_\mathrm{ff}(r,t)}{t} \right)
        \ud \theta \ud r,
\end{align}

where $\tau_\mathrm{ff}(r,t)$ is the free-fall time at a given radius at time $t$, which changes as the structure of the star evolves. For brevity, the integrand of Equation~\ref{eqn:aetwo} is represented by $\mathcal{I}$. Examples of the exponential suppression term are shown in the RHS panel of Figure~\ref{fig:amplitude} at two reference times. As time progresses, the region that is strongly suppressed moves outwards in radius until, eventually, no further suppression is applied anywhere on the grid. This reflects the expectation that the artificial signal will decrease with time as the outer layers of the star relax closer to an equilibrium state. The free-fall time is a convenient timescale for these changes to occur over, however we stress that this is an approximate and somewhat heuristic method. The result of this suppression term is to zero-out $A^{\mathrm{E2}}_{20}$ initially, and then reduce the maximum (negative) amplitude to about one third of the full signal. This approach also reduces the height of the crest by about half, but has minimal effect on the trough. 

The blue line (labeled `truncated') shows the result of simply cutting out the negative part of the waveform. To match more smoothly with the zero-padded signal (see Section~\ref{sec:detect}), the signal is first allowed to go slightly positive, and then tapered down to zero with an exponential decay function over a timescale of ${\sim} 0.3 \, \mathrm{s}$. This timescale is comparable with the collapse timescale of the core of free protons and neutrons, but is largely motivated by numerical convenience over physical accuracy. The impact of tapering the various waveforms to zero is discussed in more depth in Section~\ref{sec:detect}. After the blue line connects back to the full signal at $t=1 \, \mathrm{s}$ (slightly after black-hole formation), it is identical to the full signal.

\subsection{Waveform template}
\label{sec:template}

For a detection by matched filtering -- discussed further in Section~\ref{sec:detect} -- one needs to take into account the dependence of the waveform on stellar parameters in the regime of black-hole forming very massive stars. At similar levels of rotation that permit disk formation, large-scale asymmetries during the collapse are arguably close to maximum, and the GW amplitude likely scales with the mass of the collapsing core. The width of the signal is regulated by the free-fall time $\tau_\mathrm{ff}$ of the core, and hence by its average density via
$\tau_\mathrm{ff}\sim (G \bar{\rho})^{-1/2}$. 
Following the argument of \citet{shibata_26}, massive stellar cores approximately obey $M \propto s^{2}$, for entropy $s$. With some additional manipulations, recalling that massive stars are radiation-dominated, one can arrive at the following scaling relations,
\begin{equation}
    \frac{c^2R}{GM} \propto \bar{\rho}^{-1/3} M^{-2/3} \propto M^{-5/12},
\end{equation}
which immediately yield the mass dependence of the average core density, $\bar{\rho} \propto M^{-3/4}$. Hence, the free-fall time, and consequently the GW signal duration, depend weakly on the core mass as $\tau_\mathrm{ff} \propto M^{3/8} \propto f^{-1}_{\mathrm{GW}}$.
We expect that rescaling the amplitude and duration of the waveform accordingly provides a viable first-order approximation \citep{shibata_26}, \REV{i.e., a simple template can be expressed as $A_\mathrm{template}(t)=C_{1} A_{300}(C_{2} t)$ in terms of the computed waveform $A_{300}$ (e.g., from Figure~\ref{fig:amplitude}), and scaling parameters $C_{1}$ and $C_{2}$ (although $C_1$ is simply a normalization factor in the context of matched filtering, and not a true degree of freedom)}. An examination of the free-fall timescales of the cores of several \textsc{Kepler} models of super massive stars hints that the mass dependence of the signal width may be even weaker. Future multi-dimensional simulations based on detailed stellar evolution models will need to address the robustness of the signal shape and its scaling with stellar mass and rotation. Our waveform is available online\footnote{\url{https://doi.org/10.5281/zenodo.19478132}} for further analysis.

\section{Detectability}
\label{sec:detect}

\begin{figure}
    \centering
    \includegraphics[width=\linewidth]{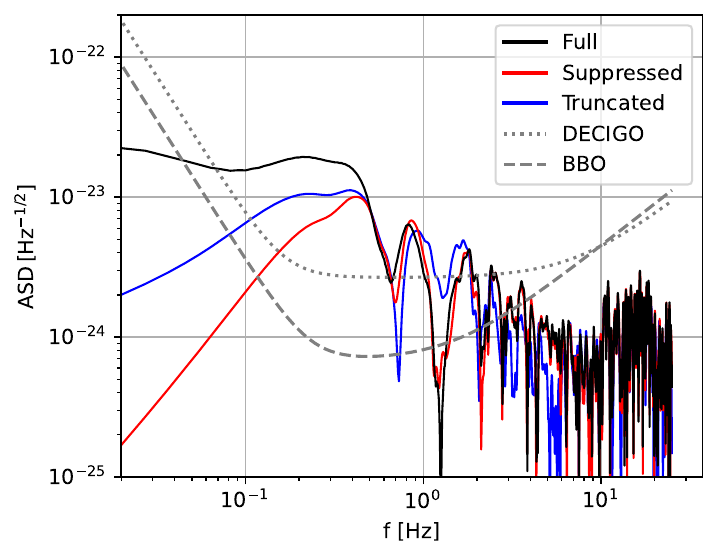}
    \caption{Amplitude spectral density (ASD) of the padded full signal (black line), and suppressed and truncated waveforms (red and blue respectively), compared to the sensitivity curves for DECIGO and BBO.}
    \label{fig:spectrogram}
\end{figure}

For determining the detectability of this signal, we zero-pad all versions of the waveform to $\pm 50 \, \mathrm{s}$. \REV{Because the full signal is significantly non-zero initially, it is tapered assuming an exponential decay of the amplitude with decay constant $1/\tau_\mathrm{ff}$, where $\tau_\mathrm{ff} \approx 0.9 \, \mathrm{s}$ is the free-fall time of the helium core. While there is some flexibility in determining the relevant timescale for the tapering of the signal, it ultimately does not impact the signal-to-noise ratio (SNR) too much. Tapering other types of low-frequency GW signals may require more finesse, however \citep{Richardson_et_al:2022}. Sections of the tapered and padded waveforms are shown in Figure~\ref{fig:amplitude}.}

Spectra of the padded signals are shown in Figure~\ref{fig:spectrogram} and compared to tentative sensitivity curves from \citet{Yagi_Seto:2011} of DECIGO \citep{DECIGO:2021} and BBO \citep{Crowder_Cornish:2005} for an event at a distance of $100\,\mathrm{Mpc}$. \REV{The spectra of all three waveforms (full, suppressed, and truncated) agree on a dominant frequency component at a few deci-Hz. The choice of signal post-processing (including tapering) mostly impacts the low-frequency tail below the peak sensitivity band of DECIGO and BBO.}

If, as motivated previously, the underlying dynamics generating the low-frequency signal are the same across moderately varying very massive progenitors, the signal morphology -- a pronounced crest and trough -- are likely similarly robust. Thus, the signal is amenable to a templated search, and we compute the signal-to-noise ratio for matched filtering following \citet{Flanagan_Hughes:1998},
\begin{equation}
    (\mathrm{SNR})^{2} = 4 \int_{0}^{\infty} \frac{|\tilde{h}(f)|^{2}}{S_h(f)} \ud f.
\end{equation}
For the noise curves used in Figure~\ref{fig:spectrogram}, the signal-to-noise ratios for an event at $100\,\mathrm{Mpc}$ are given in Table~\ref{tab:detectability}.

\begin{table}[]
    \centering
    \begin{tabular}{ccccc}
    \hline
    \hline
          & \multicolumn{2}{c}{$\mathrm{SNR_{100}}$} & \multicolumn{2}{c}{$\mathrm{D_\mathrm{SNR7}}$} \\
         Waveform & DECIGO & BBO & DECIGO & BBO\\
         \hline
         Full & 7.0 & 22.5 & 0.15 & 5.0\\
         Suppressed & 3.4 & 11.8 & 0.02 & 0.72\\
         Truncated & 4.4 & 14.6 & 0.04 & 1.36\\
         \hline
         \hline
    \end{tabular}
    \caption{Summary of detection statistics for the three waveform variants in DECIGO and BBO. The $\mathrm{SNR_{100}}$ columns give the optimally-orientated SNR for the waveforms at a distance of $100 \, \mathrm{Mpc}$. $D_\mathrm{SNR7}$ represents the number of predicted detections over a 5 year mission assuming a minimum detection threshold SNR of $7$.}
    \label{tab:detectability}
\end{table}

\REV{These SNRs indicate that BBO would have a plausible chance of detecting the collapse of a rotating very massive star; failing that, a non-detection could still provide an upper bound on the local collapse rate of such stars. To quantify this argument, we begin with a simple calculation of the detectability of the signal. This is followed by more rigorous analyses in Sections~\ref{sec:vary_params} and \ref{sec:BayesIMF}.}

Since the sensitivity of BBO and DECIGO is expected to be nearly isotropic, we can obtain the effective
detection volume simply by accounting for the dependence of the
SNR and strain $h$ on the observer angle $\theta$ with respect to the source's rotation axis, $\mathrm{SNR}\propto h\propto \sin^2\theta$. In terms of the optimal SNR at the fiducial reference distance $D=100\,\mathrm{Mpc}$, the effective detection volume $V$ for a minimum signal-to-noise ratio $\text{SNR}_\mathrm{min}$ is then,
\begin{align}
V&=
     \frac{2\pi}{3} \int_0^{\pi } \left(\frac{D \,\text{SNR}(D) \sin ^2(\theta )}{\text{SNR}_\mathrm{min}}\right)^3 \sin\theta \,d\theta 
     \nonumber
     \\
     &= \frac{64 \pi}{105} \left(\frac{\mathrm{SNR(100\, \, \mathrm{Mpc}}) \times  100\,\mathrm{Mpc}}{\mathrm{SNR_\mathrm{min}}}\right)^{3}
     .
\end{align}

The expected volumetric number of events depends on the rate $r_\mathrm{c}$ of collapse events of all massive stars above the minimum
supernova mass of ${\sim} 8\msun$, the fraction $f_\mathrm{PBH}$
of these stars in the initial mass range for prompt collapse of rotating stars to black holes above the PISN range, and the fraction $f_\mathrm{LM}$ of  stars that rotate rapidly and have sufficiently low metallicity to evolve to the pair-instability regime due to reduced mass loss. Assuming a Salpeter IMF
with power-law index $-2.35$ and a candidate mass range of
$260\texttt{-}320\msun$ yields $f_\mathrm{PBH}=2.2\times10^{-3}$. Noting that about $10\%$ of CCSN host galaxies have metallicity lower than the Large Magellanic Cloud \citep{Anderson_et_al:2015}, and that CCSNe may be biased towards lower metallicities than their host \citep{Pessi_et_al:2023}, we use $f_\mathrm{LM}=0.1$. With a volumetric supernova rate of \REV{$r_\mathrm{c} = 0.7\times10^{-4} \, \mathrm{yr^{-1} \, Mpc^{-3}}$} \citep{Pessi_et_al:2025}, this results in \REV{${\sim}1.4$ expected events with BBO during a prospective 5-year mission using the SNR of the truncated signal and assuming a baseline minimum SNR of $7$; this is discussed further in Appendix~\ref{app:minsnr}. Addition rate estimates are given in Table~\ref{tab:detectability} for the other two waveforms, and for DECIGO.}

\REV{The full and suppressed waveforms waveforms, likely being less realistic than the truncated waveform by over- and under-estimating the SNR, respectively, indicate that BBO may detect anywhere from zero to five collapses of a rotating very massive stars over a 5-year mission. In practice, this range could be broader due to other uncertainties.} For instance, $f_\mathrm{LM}$ and $f_\mathrm{PBH}$ are not well constrained due to an incomplete understanding of metallicity-dependent mass loss. The impact of stellar rotation is not fully understood. Uncertainties in the slope of the IMF (which may in turn depend on metallicity and other environmental factors \citep{tanvir_24}), and a potential high-mass cut-off of the IMF \citep{Elmegreen:2000} may also impact the predicted range of predicted events. Even a non-detection would, however, provide at least weak rate constraints, and population-level statistics may still be possible (see Section~\ref{sec:BayesIMF}).

\subsection{Varying detection rate estimate assumptions}
\label{sec:vary_params}

Numerous assumptions shape the predicted number of detections for a given waveform in Table~\ref{tab:detectability}. In this section, we vary these parameters to assess their impact on the detectability of the signal. We will continue to use the truncated waveform as it is probably the most realistic. We motivate these parameter variations below, and give predicted detection counts in Table~\ref{tab:rate_variations}:

\textbf{High-mass IMF}: The IMF of very massive stars is not well constrained either theoretically or observationally. We take two additional power-law IMFs and extrapolate them to the regime of very massive stars. Namely, we use the steeper IMF of \citet{Weisz_et_al:2015}, and the shallower one of \citet{schneider_18}, with power-law indices $-2.45$ and $-1.9$ respectively. A shallower IMF yields a larger fraction of stars at high mass, increasing the number of collapse candidates and consequently increasing the estimated detection rate.

\textbf{Candidate mass range}: As relatively few simulations of the collapse of very massive stars are available, it is not possible to rigorously determine the candidate mass range over which the collapse dynamics are similar enough to produce the same signal morphology. At the lower end, the estimate is impacted by uncertainties in the upper cutoff for PISN, while at the higher end there is simply a lack of data; thus a conservative estimate of $320 \msun$ is most appropriate. However, if the range is wider than this, additional stars become candidates in the rate estimation, increasing the predicted number of detections.

\textbf{Mission duration}: We set the default mission duration as 5 years. This is similar to that of the LISA mission \citep{Colpi_et_al:2024}. The proposed mission for DECIGO is $>3 \, \mathrm{yr}$ \citep{Kawamura_et_al:2021}, which we take as a lower bound in our calculations. An upper bound of $10 \, \mathrm{yr}$ is optimistic, but interesting nonetheless. Due to the dependence of $\mathrm{SNR_{min}}$ on the mission duration, $D_{3\sigma}$ in Table~\ref{tab:rate_variations} scales slightly nonlinearly, with a very weak `penalty' for longer observational periods.

\begin{table}[]
    \centering
    \begin{tabular}{cccc}
        \hline
        \hline
          & \multirow{2}{*}{Variant} & \multicolumn{2}{c}{$\mathrm{D}_{3\sigma}$} \\
          & & DECIGO & BBO \\
        \hline
         Baseline & & 0.04 & 1.3 \\ 
        \hline
        \multirow{2}{*}{IMF} & $-2.45$ \citep{Weisz_et_al:2015} & 0.03 & 1.0 \\ 
          & $-1.9$ \citep{schneider_18}& 0.12 & 4.4\\
        \hline
        \multirow{3}{*}{Candidate mass} & $250-320$ & 0.04 & 1.6 \\ 
          & $260-500$ & 0.09 & 3.2\\
          & $250-500$ & 0.09 & 3.5\\
          & $260-1\mathord,000$ & 0.12 & 4.5\\
        \hline
        \multirow{2}{*}{Mission duration} & $3$ & 0.02 & 0.82 \\ 
          & $10$ & 0.07 & 2.6\\
        \hline
        Optimistic & & 0.97 & 35\\
        Pessimistic & & 0.02 & 0.62 \\
        \hline
        \hline
    \end{tabular}
    \caption{Number of detections given changes in the assumptions of the rate estimate. This is expressed as $\mathrm{D}_{3\sigma}$, or detections at a significance of at least $3\sigma$. Note that this is subtly different to $\mathrm{D}_\mathrm{SNR7}$ in Table~\ref{tab:detectability} as $\mathrm{SNR_{min}}$ is not \emph{exactly} $7$ given our baseline parameters; changing the mission duration will also change $\mathrm{SNR_{min}}$ and so using $\mathrm{D}_{3\sigma}$ as the detection metric allows the most meaningful comparison. The baseline parameters are as in Section~\ref{sec:detect}: the Salpeter IMF with index $-2.35$, a candidate mass range of $260-320 \msun$, and a 5-year mission duration. We further assume a constant $10 \, \mathrm{Hz}$ sampling frequency, and disregard the (small) impact of frequency variations of the template over the range of candidate masses on the number of required samples when calculating $\mathrm{SNR_{min}}$ with Equation~\ref{eqn:snrmin}. Optimistic/pessimistic variants combine all variants which increase/decrease the number of detections.}
    \label{tab:rate_variations}
\end{table}

Even for the most optimistic combination of parameter variations, DECIGO is unlikely to detect the collapse of a very massive star. The situation is different for BBO, however, with several detections becoming likely with favorable changes to the rate calculation. The range of $0.62$ to $35$ detections between worst and best-case is quite large, reinforcing the need for more sophisticated predictions of this nature to better inform the design of future deci-Hz GW detectors.

\subsection{Constraining the high mass IMF}
\label{sec:BayesIMF}

Our preceding analysis indicates that reliable single-source detections of GWs from collapsing very massive stars may be few in number. Despite that, examining the population as a whole by also considering non-detections and lower-SNR events could provide a method of distinguishing between the astrophysical scenarios outlined in Table~\ref{tab:rate_variations}. For that purpose, we perform a Monte Carlo analysis to determine if the IMF of very massive stars can be constrained by GW observations in the deci-Hz band. 

We start with our standard detection rate parameters: a candidate mass range of $260-320 \msun$, and a mission duration of 5 years. Using the SNR of the truncated signal, $14.6$, and the IMFs with index $-2.35$ and $-1.9$, we estimate the event rate within spherical shells of detection volume, i.e., 

\begin{equation}
    \ud D_\mathrm{SNR7} = r_\mathrm{c} f_\mathrm{PBH} f_\mathrm{LM}T_\mathrm{mission} \ud V,
\end{equation}
where the terms are as defined in Section~\ref{sec:detect}, and $T_\mathrm{mission}=5 \, \mathrm{yr}$.

We then perform Monte Carlo simulations of events at each radius; $10^{7}$ simulations are sufficient to achieve good convergence. The event count in each shell is sampled from a Poisson distribution. The SNR at that distance is calculated, and then Gaussian noise is added to it. Events where the SNR remains above $\mathrm{SNR_{min}}$ count as true positives for a given simulation. The corresponding false positive rate is again drawn from a Poisson distribution with mean calculated as $N_\mathrm{samples} \times \mathrm{erfc}\left( \mathrm{SNR_{min}} / \sqrt{2} \right)$, i.e., the probability for a sample of noise to be above $\mathrm{SNR_{min}}$ multiplied by the number of samples.

\begin{figure}
    \centering
    \includegraphics[width=\linewidth]{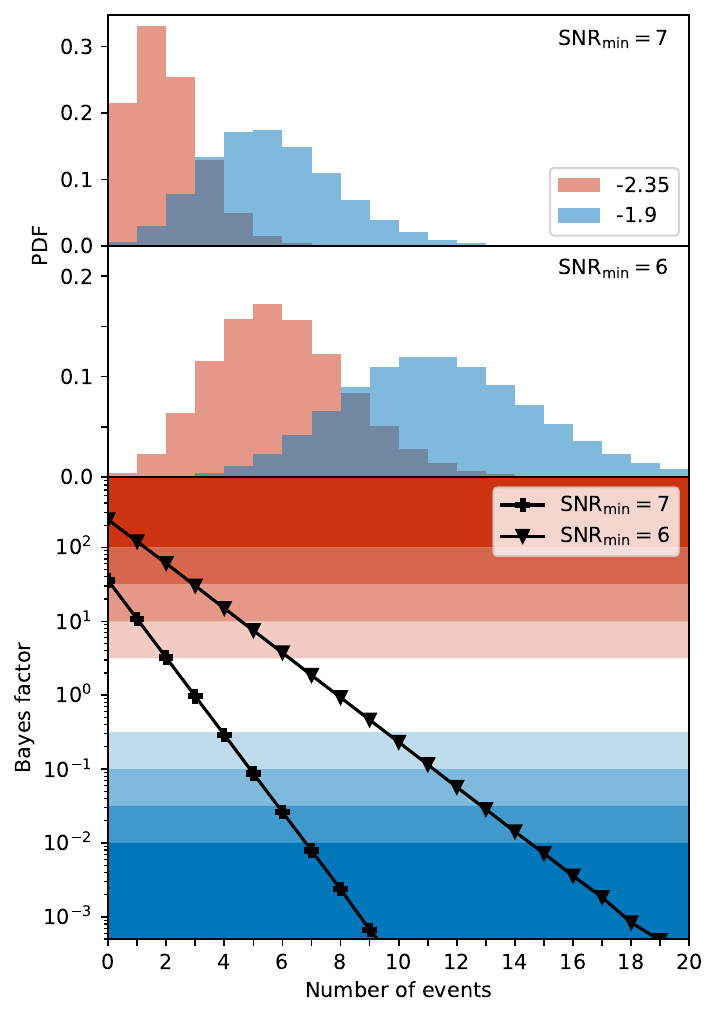}
    \caption{\textbf{Top}: probability density functions (PDFs) of simulated event counts for two IMF slopes assuming $\mathrm{SNR_{min}}$ of 7, as in Section~\ref{sec:detect}, or 6, which has too large a false positive rate for single-source detections but can be useful for population-level statistics. \textbf{Bottom}: Bayes factors comparing the two IMF slopes for the two choices of $\mathrm{SNR_{min}}$. Values $>1$ support the shallower $-2.35$ index Salpeter IMF while values $<1$ support the $-1.9$ index IMF of \citet{schneider_18}. The graduated color bands indicate the interpretation bands of the Bayes factor \citep{Jeffreys_1961}. }
    \label{fig:bayes_imf}
\end{figure}

Histograms of simulated events are shown in the top two panels of Figure~\ref{fig:bayes_imf} for $\mathrm{SNR_{min}}$ of 6 and 7. The effect of lowering $\mathrm{SNR_{min}}$ is to push the distributions upwards in event count and become broader. While the true positive rate is dependent on the IMF, the false positive rate is not, instead just scaling with $\mathrm{SNR_{min}}$. For $\mathrm{SNR_{min}}=7$ the average number of false positives is approximately $4 \times 10^{-3}$, while for $\mathrm{SNR_{min}}=6$, it is about $3$. The latter is a relatively large number of false positives, however the number of true positives is also larger, which keeps the two distributions from overlapping too much.

In fact, one can exploit this property of the distributions in the face of a high false positive rate to bolster the evidence in hypothesis testing. Assuming flat priors, the Bayes factor is simply the odds ratio of the two IMFs at given event count, $P_{-1.9}(N_\mathrm{events})/P_{-2.35}(N_\mathrm{events})$. These are plotted in the lower panel of Figure~\ref{fig:bayes_imf} along with colored bands \citep{Jeffreys_1961} indicating how strongly the evidence supports the steeper $-2.35$ index IMF (red), or shallower $-1.9$ index IMF \citep{schneider_18} (blue). 

As an example, if no events are detected, both choices for $\mathrm{SNR_\mathrm{min}}$ support the Salpeter (-2.35) IMF hypothesis; for $\mathrm{SNR_{min}}=7$, the support is `very strong', while for $\mathrm{SNR_{min}}=6$ the support is classified as `decisive'. On the other hand, at 7 events, the larger $\mathrm{SNR_{min}}$ indicates that is `decisive' support for the IMF of \citet{schneider_18}, whereas achieving this result with the smaller $\mathrm{SNR_{min}}$ does not provide conclusive evidence either way. Detecting 5 events provides a similar level of support for either hypothesis depending on the choice of $\mathrm{SNR_{min}}$. We stress that this does \emph{not} mean that the preferred IMF model can change depending on the threshold SNR used -- the number of events will increase/decrease as the threshold is decreased/increased. Instead, model selection simply makes a stronger statement in some cases depending on $\mathrm{SNR_{min}}$, e.g., if no events are detected at a lower minimum SNR.

\section{Conclusions}
\label{sec:conclusions}

\REV{Our simulations suggest that the collapse of very massive stars just above the PISN range will be a challenging, although potentially viable target for future space-based GW detectors in the deci-Hz range. The most ambitious these, BBO, is estimated to detect a few such events over a five-year mission.}  Our simulations add another scenario in the landscape of GW emission from high-mass stellar sources.

Similar to the collapse of supermassive stars \citep{montero_12,shibata_16,uchida_17}, we find an interesting signal even without the onset of triaxial instabilities during the collapse. We note, however, that the dynamics of the collapse of very massive stars is significantly different to supermassive stars as it is triggered by a different instability (i.e. the pair instability), starts from a different evolved stellar structure, and is influenced by nuclear burning in a very different manner. The collapse of very massive stars is also much less speculative, \REV{such that a tentative rate estimate is possible. This yields a result of ${\sim}0.3$ deci-Hz GW events per year with BBO.}
Unlike previous general relativistic simulations of very massive stars \citep{shibata_16, shibata_26}, the GW signal we consider does not depend on triaxial instabilities in the disk. These would contribute GWs at higher frequencies of ${\sim} 10\,\mathrm{Hz}$ and may already be detectable in current ground-based detectors out to a similar distance \citep{shibata_26}. However, 3D simulations are needed to confirm the onset and growth of these instabilities.

\REV{We have shown that our detection rate estimates are sensitive to the assumed mass range of very massive stars that collapse to BHs in a similar manner above the PISN regime. Additional simulations over a wide mass range of very massive stars are needed to confirm the uniformity of the low-frequency waveform over the full range of very massive candidates.
Even if a single-source detection with a high SNR is not realized over the mission duration, we further demonstrate that it may be possible to statistically constrain the slope of the IMF in the regime of very massive stars.}

\REV{The robustness of our predictions is primarily limited by the non-hydrostatic initial configuration of the rotating progenitor, which deforms due to centripetal forces after mapping from a 1D stellar evolution code. This large-amplitude component of the low-frequency signal should be mitigated in future work to produce more reliable waveforms for detection estimates and strengthen the science case for proposed deci-Hz detectors. }

Finally, foregrounds and backgrounds form a potential impediment to the detection of the GW signal from the collapse of rotating very massive stars.
The foreground from binary inspirals consists of resolvable events, and can therefore likely be removed \citep{Cutler_Harms:2006}.
The stochastic GW background from inflation \citep{ChiaraGuzzetti_et_al:2016} may inherently limit detectability for larger values of the spectral energy density $\Omega_\mathrm{GW}$. A detectability analysis including the background and foreground is desirable, and should also be considered in future work.

Our work underscores the opportunities for GW astronomy in the deci-Hz range, for which long-range planning is currently ongoing \citep{Perego_et_al:2025}.
Future 3D simulations at the upper end of the pair-instability regime are required to more fully investigate GW emission and the associated multi-messenger signatures, and will need to model in more detail the processes that govern disk evolution and outflows after black-hole formation.

\section*{Data availability}
The predicted GW waveform is publicly available \citep{data_waveform:2026}. Further data from our simulations will be made available upon reasonable requests made to the authors. 

\section*{Acknowledgements}
The authors acknowledge support by the Australian Research Council through grants DP240101786 (BM, AH),
LE260100008 (JP), LE230100063 (AH), and DP240103174 (AH).
The authors acknowledge support
by the Australian Research Council (ARC) Centre of Excellence (CoE) for Gravitational Wave Discovery (OzGrav) project number CE230100016. 
This project was carried out using computer time allocations from Astronomy Australia Limited's ASTAC scheme, the National Computational Merit Allocation Scheme (NCMAS), and
from an Australasian Leadership Computing Grant.
Some of this work was performed on the Gadi supercomputer with the assistance of resources and services from the National Computational Infrastructure (NCI), which is supported by the Australian Government.

\bibliographystyle{apsrev4-2-truc_auth}
\bibliography{bibliography.bib}

\appendix

\section{Minimum SNR}
\label{app:minsnr}

The SNR of noise is, by design, normally distributed. Thus, defining $P(\mathrm{FP}>0)$ as the probability of any number of false positive (FP) detections after sampling $N_\mathrm{samples}$ times, one can write,

\begin{align}
    P(\mathrm{FP}>0) & = 1 - P(\mathrm{FP}=0) \\
    & = 1 - \left[ \mathrm{erf} \left(  \frac{\mathrm{SNR_{min}}}{\sqrt{2}} \right) \right] ^{N_\mathrm{samples}},
\end{align}

where $\mathrm{SNR_{min}}$ is the threshold above which a sample counts as a detection.

In order to control the FP rate in matched filtering, this probability should be below some threshold level. It is often convenient to express this in multiples of the standard deviation of the normal distribution, e.g., $3\sigma$, while we denote the corresponding probability of being outside that range as $P_\sigma$. As such,

\begin{equation}
    1 -  \left[ \mathrm{erf} \left(  \frac{\mathrm{SNR_{min}}}{\sqrt{2}} \right) \right] ^{N_\mathrm{samples}} = P_\sigma
\end{equation}
Solving for $\mathrm{SNR_{min}}$ gives,
\begin{equation}
    \label{eqn:snrmin}
    \mathrm{SNR_{min}} = \sqrt{2} \, \mathrm{erf}^{-1}\left( \left( 1 - P_\sigma \right)^{1/N_\mathrm{samples}} \right)
\end{equation}

This threshold SNR depends on $P_\sigma$, the desired significance level, and $N_\mathrm{samples}$, the number of attempted matches during the matched filtering.

As a baseline, we use $3\sigma$ significance: $P_\sigma = 2.7 \times 10^{-3}$. Meanwhile, $N_\mathrm{samples}$ scales with the signal duration $T_\mathrm{duration}$. For signals in the deci-Hz range, it is reasonable that a $10\, \mathrm{Hz}$ sampling frequency is required to find good overlap of the signal and the template. Additionally, assuming a template of the form given in Section~\ref{sec:template} introduces additional sampling requirements. Since we predict $f_\mathrm{GW} \propto M^{-3/8}$, i.e., weak scaling of the deci-Hz GW frequency with progenitor mass, the variation in frequency over the candidate range of $260-320  \msun$ is less than $10\%$. Another factor of $10$ in $N_\mathrm{samples}$ should cover such a frequency variation adequately. This yields, $N_\mathrm{samples} = 5 \, \mathrm{yr} \times 10 \, \mathrm{Hz} \times 10 \approx 1.5 \times 10^{9} $. More complex templates naturally require additional sampling \citep{Cutler_Harms:2006}.

With these baseline $P_\sigma$ and $N_\mathrm{samples}$ set, Equation~\eqref{eqn:snrmin} gives $\mathrm{SNR_{min}} \approx 7$. This is the value used in Section~\ref{sec:detect} and Table~\ref{tab:detectability}. If one can tolerate a larger $P_\sigma = 0.05$ ($2\sigma$) then the very weak scaling of $\mathrm{SNR_{min}}$ with $N_\mathrm{samples}$ means an additional factor of ${\sim} 10^{5}$ can be included in $N_\mathrm{samples}$ for more template flexibility, while only increasing $\mathrm{SNR_{min}}$ to $8$.

\end{document}